\documentclass{svproc}

\usepackage{url}
\usepackage{nameref}
\usepackage{hyperref}

\usepackage{multirow}
\usepackage{tabularx}
\usepackage{graphicx}
\usepackage{subcaption}
\usepackage{enumitem}
\begin{document}
\mainmatter              

\title{Simulation-based Testing of Foreseeable Misuse by the Driver applicable for Highly Automated Driving }
\titlerunning{Simulation-based testing of FM by the driver applicable for HAD}  
\author{Milin Patel\inst{1} \and Rolf Jung\inst{1} \and Yasin Cakir\inst{2}}

\authorrunning{Milin Patel et al.} 
\tocauthor{Milin Patel, Yasin Cakir, Rolf Jung}

\institute{Institut für Fahrerassistenz und vernetzte Mobilität, 87734, Germany\\
	\email{\{milin.patel, rolf.jung\}@hs-kempten.de}
	\and
	University of Applied Sciences Kempten, 87435, Germany\\
	\email{yasin.cakir@stud.hs-kempten.de}}

\maketitle

\begin{abstract}
With Highly Automated Driving (HAD), the driver can engage in non-driving-related tasks. In the event of a system failure, the driver is expected to reasonably regain control of the Automated Vehicle (AV). Incorrect system understanding may provoke misuse by the driver and can lead to vehicle-level hazards. ISO 21448, referred to as the standard for Safety of the Intended Functionality (SOTIF), defines misuse as usage of the system by the driver in a way not intended by the system’s manufacturer. Foreseeable Misuse (FM) implies anticipated system misuse based on the best knowledge about the system’s design and the driver’s behavior. This is the underlying motivation to propose simulation-based testing of FM. The vital challenge is to perform a simulation-based testing for a SOTIF-related misuse scenario. Transverse Guidance Assist System (TGAS) is modeled for HAD. In the context of this publication, TGAS is referred to as the “system,” and the driver is the human operator of the system. This publication focuses on implementing the Driver-Vehicle Interface (DVI) that permits the interactions between the driver and the system. The implementation and testing of a derived misuse scenario using the driving simulator ensure reasonable usage of the system by supporting the driver with unambiguous information on system functions and states so that the driver can conveniently perceive, comprehend, and act upon the information.

\end{abstract}

\section{Introduction}

In HAD, longitudinal and lateral vehicle guidance is controlled by the system. However, when the system reaches its operational limits, the Human Driver (HD), referred to as a fallback-ready user in SAE J3016 taxonomy \cite{SAElevel4}, is expected to regain driving control within a reasonable amount of time. Whenever the system is not capable of handling a situation within its Operational Design Domain (ODD), a Take-Over Request (TOR) is issued by the system as a notification indicating that the HD should promptly perform the driving tasks. \\

Transition in Automated Driving (AD) is the process and period of transferring responsibility and driving control over some or all aspects of the driving tasks between HD and the system. Transition can be either activation or deactivation of a function or a change from one driving state to another as per \cite{ZhenjiLu.2016}. According to the SAE J3016 taxonomy \cite{SAElevel4}, the driver has no active role or driving responsibilities when the system is operating within its ODD. Engagement in non-driving related task keeps the driver out of the loop, which leads to misuse when returning to Manual Driving (MD) in take-over situations \cite{doi:10.1177/0018720812461374}. \\

\begin{figure}
	\centering
	\includegraphics[width=\linewidth]{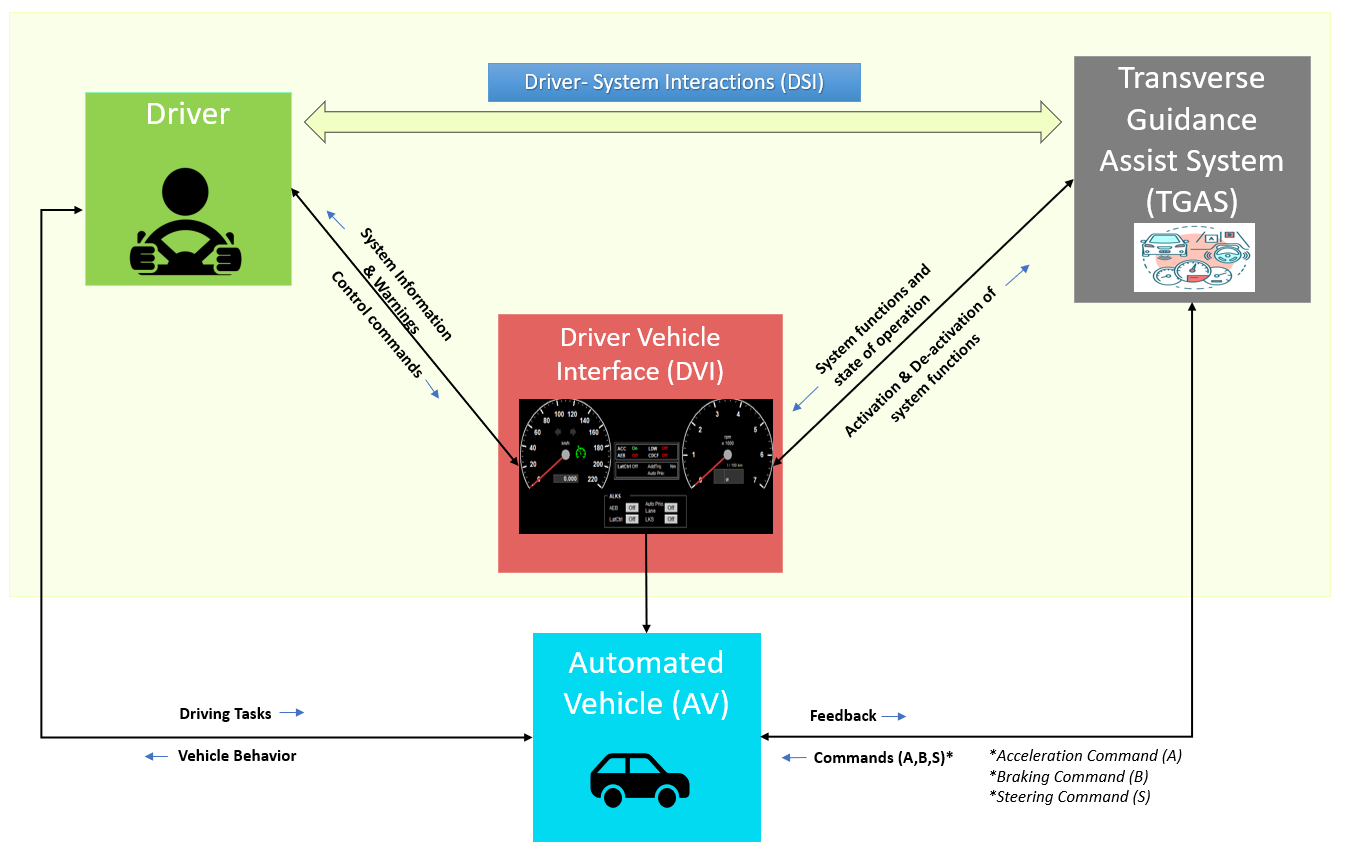}
	\caption{Incorporation of the driver with the system and AV: interactions and the interface}
	\label{fig1}
\end{figure}

To ensure smooth transition from AD to MD, the TOR must be presented through a well-designed interface. Therefore, the implications of Driver-Vehicle Interface (DVI) design on the interactions between driver and the system, abbreviated as Driver-System Interactions (DSI), must be studied so the driver can regain control over HAD while deterring misuse. The Figure ~\ref{fig1} depicts a pictorial representation of the incorporation of the driver in terms of interactions with the system and interface with the Automated Vehicle (AV). One of the key subject in the SOTIF standard is FM, which is substantial consideration for human factors engineering \cite{Walker.2019b}. It should be noted that this publication focuses on FM by the driver and as part of the testing, human factors during transition in HAD are taken into consideration, not the other way around.  

The factors for FM considered in this publication are driver’s Recognition \& Judgment. Simultaneously, False recognition \& Misjudgment, by the driver are causes for FM. The aforementioned factors and causes addressed in this publication are referred to as, human misuse process and guidewords, in the informative annex B of ISO 21448 \cite{ISO_21448}. The False recognition is analogous to the perceptual errors, where the driver’s perception of the environment differs from reality. The Misjudgment is akin to the decision errors, in which the driver decides on an incorrect action for the given situation. \cite{NHTSA_SOTIF}\\ 

This publication is structured as follows. Section \ref{Chap_2} refers to description of a SOTIF-related misuse scenario. Section \ref{Chap_3} outlines the strategy for implementing simulation-based testing of FM. Section \ref{Chap_4} addresses the implementation using driving simulator and elaborates on results. Finally, Section \ref{Chap_5} presents concluding remarks.

\section{SOTIF-related misuse scenario}
\label{Chap_2}
SOTIF-related misuse scenario can be derived from gained knowledge and brainstorming \cite{ISO_21448}. The methodology for systematically deriving SOTIF-related misuse scenario to support the safety analysis for the system is provided in the ISO 21448 \cite{ISO_21448}.  
\begin{figure}[htbp]
	\centering
	\includegraphics[width=\linewidth]{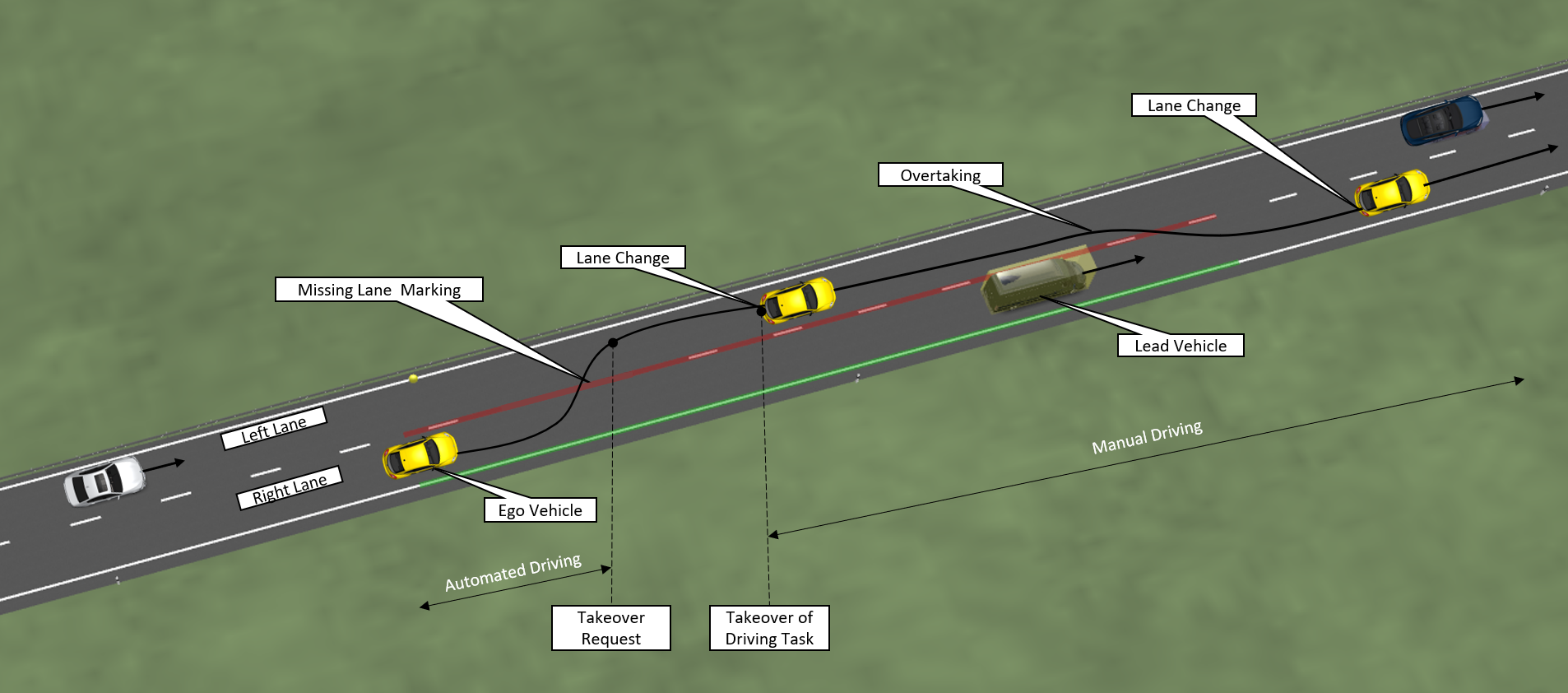}
	\caption{Highway Lane Change Scenario}
	\label{fig2}
\end{figure}
The scenario shown in Figure ~\ref{fig2} is derived to expose the driver to a situation requiring transverse guidance that is formed by lateral and longitudinal maneuvering of the Ego-Vehicle in a high-speed highway environment. Ego-Vehicle in the context of this publication is defined as the AV equipped with TGAS. The entire scenario is divided into three maneuvers:

\begin{itemize}
	\item Lane change from right to left lane
	\item Overtaking of lead-vehicle from left lane
	\item Lane change from left to right lane
\end{itemize}

The Table~\ref{scenariotable} outlines the derived SOTIF-realted misuse scenario considered in this publication conforming to an example methodology given in annex B of ISO 21448 \cite{ISO_21448}. 
\begin{table*}[h]
	\caption{Description of SOTIF-related misuse scenario}
	\label{scenariotable}
	\begin{center}
		\centering
		\begin{tabularx}{\textwidth}{|X|X|X|X|}
			\hline
			\centering
			\multirow{2}{*}{\textbf{Stake-holder}}                                                    
			&  \multicolumn{2}{c|}{\textbf{Foreseeable Misuse (FM)}}                             
			& 
			\multirow{2}{*}{\textbf{\begin{tabular}[c]{@{}c@{}}Driver System \\Interactions (DSI)\end{tabular}} } 
			\\ \cline{2-3}
			&                                                                    \centering \textbf{\begin{tabular}[c]{@{}c@{}}Factors\end{tabular}} 
			&  \centering \textbf{\begin{tabular}[c]{@{}c@{}}Causes\end{tabular}} &                                                                                   \\ \hline
			\centering	\multirow{3}{*}{\textbf{Driver}}                  
			&
			\centering Recognition 
			& 
			
			\centering False recognition
			& Delayed Take-over
			
			\\ \cline{2-4}
			&
			\centering Judgement 
			& 
			
			\centering Misjudgement
	
			&  \begin{tabular}[c]{@{}c@{}}Take-over \& perform\\Under-steer\end{tabular}
			
			\\ \hline

		\end{tabularx}

		\begin{tabularx}{\textwidth}{|X|X|X|}
	
			\hline
			\centering
			\textbf{\begin{tabular}[c]{@{}c@{}}Environmental \\conditions\end{tabular}}
		
			&
			\centering
			\textbf{\begin{tabular}[c]{@{}c@{}}Potential SOTIF-\\related misuse scenario \end{tabular}}
		
			& 
			
			\textbf{\begin{tabular}[c]{@{}c@{}}Derived hazardous \\misuse scenario\end{tabular}} 
		
			\\ \hline
			
			\centering 	\begin{itemize} 	\item Weather : clear  	\item Light Condition : daylight 	\item Traffic Condition : light traffic 	\item Roadway Surface and Features : missing lane markings 	\end{itemize} &
			The Ego-Vehicle encounters a road with missing lane markings during automated driving on a two lane one-way highway and executing lane change maneuver from right to left lane.

			The camera sensor cannot estimate the location of the lane boundary due to a performance insufficiency. Ego-Vehicle starts to leave the lane and driver is notified to take control of the driving tasks by means of TOR.
			
			& 
		
			Driver fails to take-over the control of the driving tasks, resulting in lane departure of Ego-Vehicle.
			
			\\ \hline
		\end{tabularx}	
	\end{center}
\end{table*}
 The DSI that influence the vehicle-level hazard related to FM, namely lane departure, are considered for deriving a SOTIF-related misuse scenario. Take-over is defined as transfer of the driving control between Human Driver (HD) and the system \cite{ISO.2020}. Under-steer means the driver fails to provide adequate steering input for the Ego-Vehicle to follow the lane.

\section{Simulation-based Testing}
\label{Chap_3}
\begin{figure}[htbp]
	\centering
	\includegraphics[width=\linewidth]{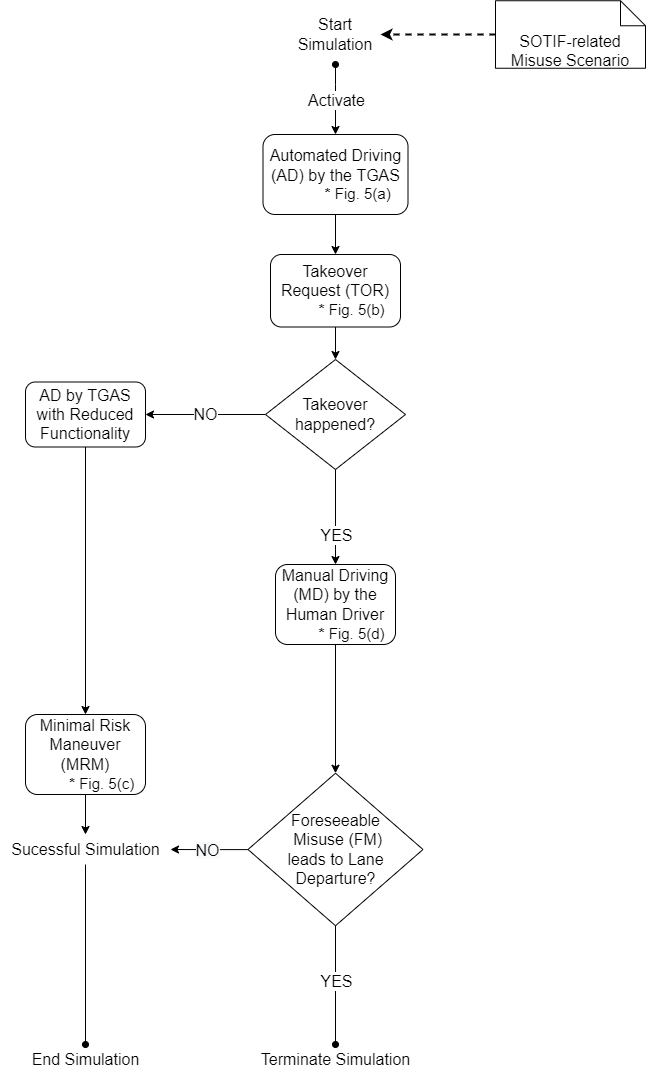}
	\caption{Simulation-based testing}
	\label{fig3}
\end{figure}

The strategy depicted in the Figure ~\ref{fig3} describes the steps in methodical order to perform simulation-based testing of the SOTIF-related misuse scenario described 

in Table ~\ref{scenariotable}. The scenario and maneuvers is modeled using IPG CarMaker, as per the description given in Section \ref{Chap_2}. The TGAS performs AD of the Ego-Vehicle by providing lateral and longitudinal control in the modeled scenario. When the system reaches its operational limits, it notify the driver by issuing TOR. Driver take-over at operational limits and the corresponding TOR are not obligatory in HAD \cite{MRM_ROI}. The system is expected to remain operational until the driver is able to regain control \cite{NHTSA_SOTIF}.\\

The driver might not be able to take-over driving control within a specified take-over time and FM is expected, attributing false competencies to the system. It may lead to lane departure of the vehicle, addressed as a vehicle-level hazard. If the driver does not take-over the driving tasks in the event of TOR, the system will transition to the AD with reduced functionality. Subsequently, a minimal risk maneuver (MRM) \cite{safetyfirst} is performed by the system to keep the Ego-Vehicle in its lane and to automatically stop the Ego-Vehicle on the side of the road in a safe manner \cite{Docsroom.2022}. The driver may be asked to take-over at the end of the MRM.
\section{Implementation and Results}
\label{Chap_4}
\begin{figure}[htbp]
	\centering
	\includegraphics[scale=0.45]{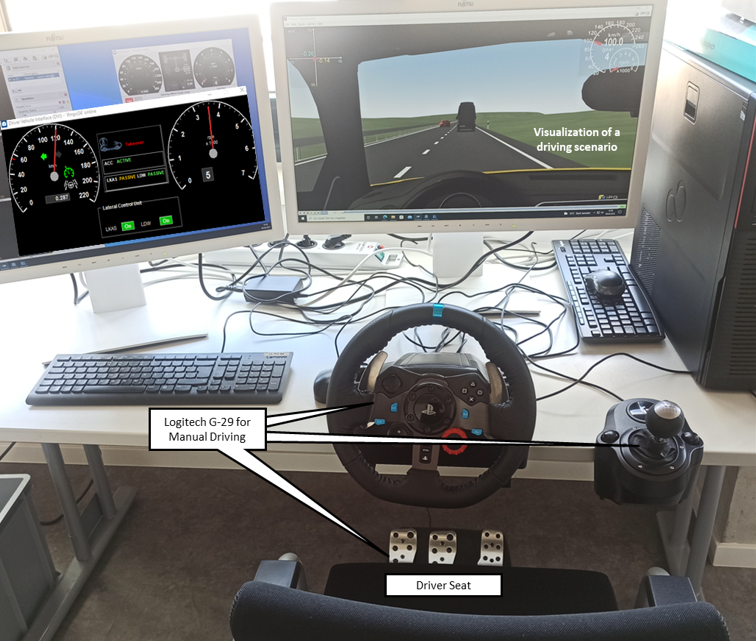}
	\caption{Driving Simulator}
	\label{fig4}
\end{figure}

The simulation-based testing is carried out using a driving simulator as illustrated in Figure ~\ref{fig4} allowing driver to control the Ego-Vehicle in the virtual test environment.

\begin{figure*}
	\centering
	\begin{subfigure}[b]{0.475\textwidth}
		\centering
		\includegraphics[width=\textwidth]{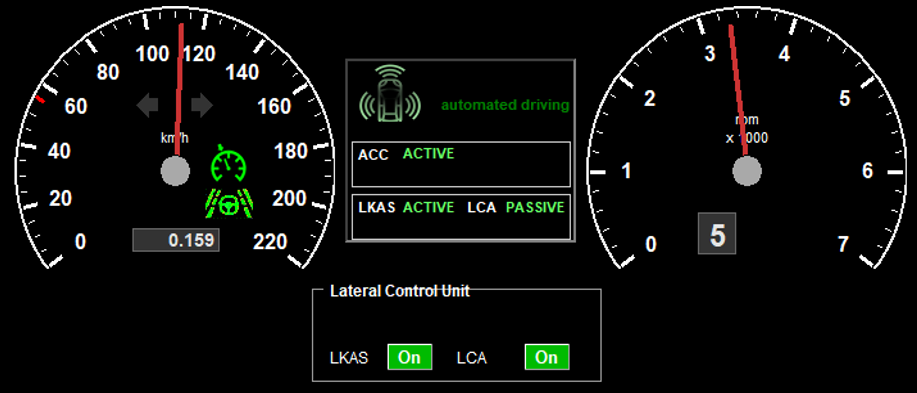}
		\caption[]%
		{{\small Automated Driving (AD)}}    
		\label{fig5a}
	\end{subfigure}
	\hfill
	\begin{subfigure}[b]{0.475\textwidth}  
		\centering 
		\includegraphics[width=\textwidth]{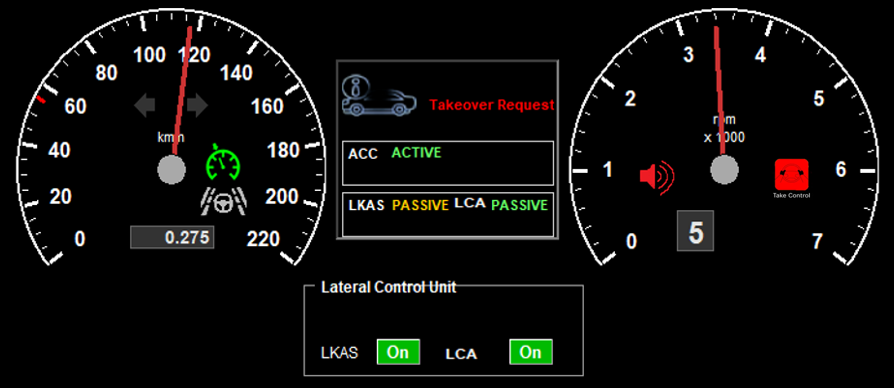}
		\caption[]%
		{{\small Take-Over Request (TOR)}}    
		\label{fig5b}
	\end{subfigure}
	\vskip\baselineskip
	\begin{subfigure}[b]{0.475\textwidth}   
		\centering 
		\includegraphics[width=\textwidth]{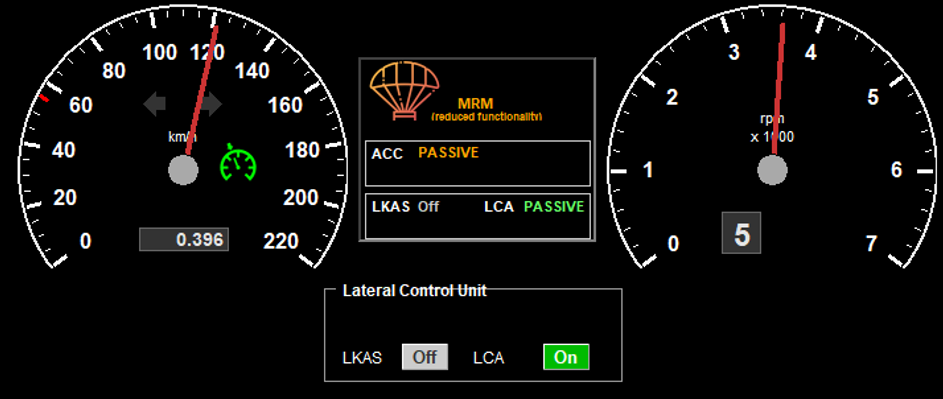}
		\caption[]%
		{{\small Minimal Risk Maneuver (MRM)}}    
		\label{fig5c}
	\end{subfigure}
	\hfill
	\begin{subfigure}[b]{0.475\textwidth}   
		\centering 
		\includegraphics[width=\textwidth]{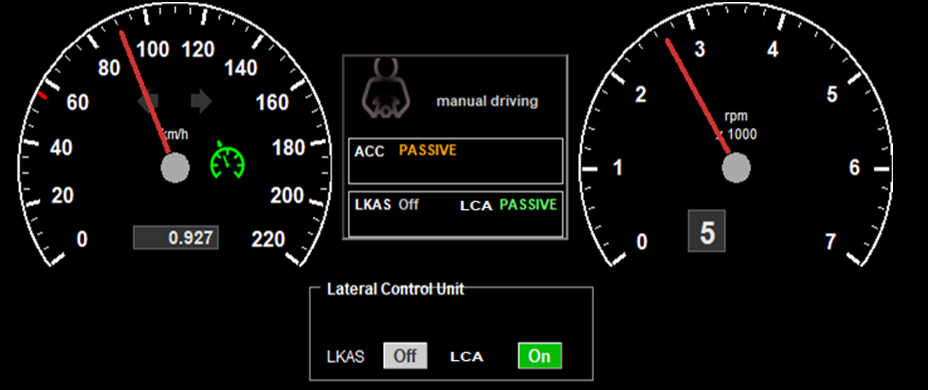}
		\caption[]%
		{{\small Manual Driving (MD)}}    
		\label{fig5d}
	\end{subfigure}
	\caption[ ]
	{\small Implementation of Driver Vehicle Interface (DVI)} 
	\label{fig5}
\end{figure*}

The driving simulator is equipped with the hardware tools (Logitech G29- steering wheel, pedals, and gearbox) integrated with a simulation tool (IPG CarMaker). Likelihood that the driver can cope with the driving situation including operational limits and system failures is determined using the driving simulator. The Driver Vehicle Interface (DVI) as illustrated in the Figure ~\ref{fig5} is designed to incorporate the interactions between driver and the system. The DVI design is in concordant with the design guidance provided in \cite{dvidesign}. 

Based on literature study pertaining to the design of automated take-over requests in HAD from diverse aspects such as procedure \cite{procedure}, timing \cite{time.takeover} and modality \cite{modality}, the TOR is cued by a combination of an auditory alert and visual notification on the designed DVI. The HD does the take-over of driving control by pushing a button on the steering wheel of driving simulator. It is conceivable that the HD might engage in FM, especially if the HD is convinced that the HAD operates practically flawless \cite{Fuest.2020} and will prevent vehicle-level hazard in the driving environment by choosing safe driving maneuvers \cite{Josten.2018}. \\

A limitation of the current implementation is the usage of a static driving simulator, where haptic motion cannot be experienced. However, the implemented DVI make it easier to keep the driver’s workload at an acceptable level by providing simultaneous auditory alert and visual notification to the driver with supporting information about the system functions and state of operation. 
\section{Conclusion and Future Work}
\label{Chap_5}

When driving with Highly Automated Driving (HAD), the Human Driver(HD) is allowed to engage in non-driving related tasks. It is reasonable to anticipate higher likelihood of the system misuse \cite{NadjaSchomig.2015}. The publication outlines the concept of Foreseeable Misuse (FM) emphasized in SOTIF standard \cite{ISO_21448} for a described SOTIF-related misuse scenario, illustrated in Section \ref{Chap_2}, applicable to HAD. An exemplary of the strategy defined for implementing simulation-based testing of FM resulting from system-initiated transition between HD and the system is demonstrated in Section \ref{Chap_3}. It should be noted that the implementation shown in Section \ref{Chap_4} is intended to demonstrate an approach for simulation-based testing of FM rather than to be a distinctive or optimal measure dedicated to mitigate FM. The relevance of this publication is that it adds to the understanding of the factors and causes contributing to FM by incorporating the concepts of DVI \& DSI, and applies it to a SOTIF-related misuse scenario. \\ 

The fundamental premise is to incorporate and manage all driver and system interactions. The simulation-based testing approach is applied to investigate the factors and mitigate the causes responsible for FM by the driver. The incorporation of DSI and DVI to address FM is briefly described but has not been evaluated. A reasonable next step for future work will be to characterize and quantify the DSI, considering aspects of HD take-over in HAD. Analysing the system specification for inappropriate interactions by the driver is a brainstorming task. One of the possible approach for analysis is System-Theoretic Process Analysis (STPA) which aims to identify the hazardous interactions in the absence of system failures \cite{BernhardKaiser.2019}. Identification of factors contributing FM by STPA and effectiveness of the mitigation measures to prevent FM for the described SOTIF-related misuse scenario are  suggested for future work. The implication of the proposed method is to exhibit how the concepts of DVI and DSI are interrelated with FM. Recommendations can be made on how, DVI design, TOR presentation modalities, driver improper interactions with system, can be adopted to address the challenges of risks that could impact functionality of HAD.

\bibliographystyle{IEEEtran}

\bibliography{references}

\end{document}